\documentclass[pdflatex,sn-nature]{sn-jnl}
\usepackage{graphicx}%
\usepackage{multirow}%
\usepackage{amsmath,amssymb,amsfonts}%
\usepackage{amsthm}%
\usepackage{mathrsfs}%
\usepackage[title]{appendix}%
\usepackage{xcolor}%
\usepackage{textcomp}%
\usepackage{manyfoot}%
\usepackage{booktabs}%
\usepackage{algorithm}%
\usepackage{algorithmicx}%
\usepackage{algpseudocode}%
\usepackage{listings}%
\usepackage{multirow}%
\usepackage{rotating}
\usepackage{soul}
\usepackage{ulem}
\sethlcolor{yellow}
\usepackage{lineno}
\begin{document}
\title[Article Title]{Negative thermal expansion in ice I polytypes}
\author*[1]{\fnm{Leonardo} \sur{del Rosso}}\email{l.delrosso@ifac.cnr.it}
\author[2]{\fnm{A. Dominic} \sur{Fortes}}
\author*[1]{\fnm{Daniele} \sur{Colognesi}}\email{d.colognesi@ifac.cnr.it}
\author[3,4]{\fnm{Alberto} \sur{Santonocito}}
\author[1]{\fnm{Francesco} \sur{Grazzi}}
\author[1]{\fnm{Selene} \sur{Berni}}
\author[1]{\fnm{Milva} \sur{Celli}}
%
%
\affil*[1]{\orgname{Consiglio Nazionale delle Ricerche}, \orgdiv{Istituto di Fisica Applicata `Nello Carrara'}, \orgaddress{\city{Sesto Fiorentino}, \postcode{I-50019}, \country{Italy}}}
\affil[2]{\orgdiv{ISIS Neutron and Muon Source}, \orgname{STFC Rutherford Appleton Laboratory}, \orgaddress{\city{Chilton}, \postcode{OX11 0QX}, \state{Oxfordshire}, \country{United Kingdom}}}
\affil[3]{\orgname{Università degli Studi di Pisa}, \orgdiv{Dipartimento di Chimica}, \orgaddress{\city{Pisa}, \postcode{I-56124}, \country{Italy}}}
\affil[4]{\orgname{Consiglio Nazionale delle Ricerche}, \orgdiv{Istituto Nazionale di Ottica}, \orgaddress{\city{Sesto Fiorentino}, \postcode{I-50019}, \country{Italy}}}
\abstract{The fundamental properties of ice have always attracted a lot of interest due to omnipresence of ice in many different natural contexts. Since cubic ice recently become experimentally accessible from a low-density gas hydrate precursor \cite{delRosso20, Komatsu20}, it has been possible to measure its density as a function of temperature in the whole thermodynamic range of metastability. We found strong analogies with respect to the other ice I polytype, i.e., hexagonal ice Ih \cite{Fortes18}, including the presence of a negative thermal expansion behavior at low temperature. Based on these results, a new enthalpy calculation quantifies the metastable nature of the cubic form and, consequently its inaccessibility from a `normal' ice Ih precursor.}
\maketitle
\newpage

\section{Main}
\label{Main}

Behind the terms `normal ice', commonly used to indicate the ice form stable at standard conditions, it hides a surprising complexity, continuously investigated by the scientific community due to the key role played by the water in fundamental physics and chemistry, as well as in many different contexts ranging from biology to astrophysics \cite{Petrenko99}. The rich polymorphism (and polytypism) presented by the crystalline-ice phase diagram, together with the development of advanced high pressure techniques, push scientists towards a seemingly eternal pursuit of new ice phases, focusing experimental and computational efforts on the different peculiarities that characterize the the various ice structures. In particular, the combined use of diamond anvil cell and laser heating allowed to investigate pure ice at extreme conditions, i.e., comparable with the interiors of Uranus and Neptune (up to 150-190 GPa and 5000-6500 K), where superionic ice phases could be present \cite{Prakapenka21}.  
More recently, an emerging class of experiments on both supercompressed \cite{lee2025multiple} and supercooled water \cite{kobayashi2025new} led to the discovery of three new metastable ice polymorphs, demonstrating, once again, the structural complexity of water as widely predicted by computational works. 
In addition, the fine tuning of different strategies like both the isothermal annealing and the addition of defects (acids/bases doping and isotopic substitution) opens new perspectives in the proton-ordering of ice polymorphs \cite{Tonauer23}, with even a multiple proton-ordered phases with respect to the same proton-disordered mother phase \cite{Hansen21}. The degree of complexity is furtherly increased when ice is combined with molecules such as small hydrocarbons, noble gases, H$_2$, N$_2$, CO$_2$, or O$_2$, which can give rise to hydrates \cite{Petrenko99}. These compounds are presumably the main constituents of the first inner layers of icy planets and satellites of the Solar system (i.e. Titan, Enceladus \cite{lunine2002cassini}), and their stability in pressure and temperature is of great interest to model the formation, evolution, and the static and dynamic properties of these celestial bodies \cite{mousis2002evolutionary}. Indeed, phenomena like cryo-volcanism, subduction processes, or warming are thought to be responsible for the destabilization of these reservoirs, thus explaining for instance the abundance of methane in Titan's atmosphere compared to the scarce quantities revealed on its surface \cite{niemann2005abundances}. In this perspective, the further investigation of the metastability of the `emptied' hydrate structure led to the discovery of low-density ice phases characterized by an open structure \cite{Falenty14,delRosso16}, which, in some cases, show an unexpected porous behaviour, opening also the route to possible applications with the ice as an innovative functional material \cite{delRosso24}. 

Beside this, there is a further intriguing aspect of the ice crystalline structure to be considered, that is the stacking arrangements of the six-membered ring water molecules layers in ice I, that gives rise to three different phases, usually named polytypes. Two out of the three are stacking-ordered and are characterized by a fully hexagonal (ice Ih) or cubic (ice Ic) crystal symmetry, while the third phase is a stacking-disordered one (ice Isd) and is characterized by the  `cubicity', i.e., the relative amount of cubic stacking sequences present in the sample \cite{Kuhs12}. The ice Ih phase, also known as `normal ice', is the stable one at ambient pressure condition and is omnipresent in many natural environments. On the contrary, ice Ic, according to computational indications \cite{Engel15}, is metastable at ambient pressure and is consequently quite difficult to find in nature as well as to synthesize in laboratory, even if its existence was deduced by the Earth's atmosphere observations \cite{Riikonen00} and it has potentially been shown to play a possible role for the cloud formation at higher altitudes \cite{Shilling06,Hervig01,Lupi17}. \\
\par
As a matter of fact, for about 50 years, all the attempts to synthesize it, following different experimental paths \cite{Salzmann20}, always led to the formation of ice Isd, reaching the 80 \% of cubicity by homogeneous nucleation in deeply supercooled liquid nanodrops \cite{Amaya17}. In 2020, the discovery of some effective routes to obtain pure cubic ice \cite{delRosso20, Komatsu20}, even by heterogeneous nucleation \cite{Davies21}, opened up the possibility to directly measure the property of this elusive ice form. 

Here, using the D$_2$O ice XVII as the mother phase \cite{delRosso21bis, Mochizuki25}, we were able to recover at ambient pressure a large amount of D$_2$O cubic ice in a powder form \cite{delRosso20}. This pivotal fact makes possible the use of the high-resolution neutron powder diffractometer HRPD (ISIS, U.K.) \cite{Ibberson09} for a careful \textit{ex-situ} characterization of its structure, extracting in this way an accurate thermal behaviour of the structural parameters in the whole metastability thermodynamic range of the cubic ice. Then, the cubic ice sample was transformed into the hexagonal one by heating the measurement cell at a constant rate, in order to also measure its structural parameters in the same thermodynamic range and experimental conditions of the previous phase, allowing in this way a direct comparison of the density for the two polytypes. The details about the sample preparation, the neutron diffraction measurements and the data analysis are reported in the Methods section.\\
\par
\begin{figure}[h]
\centering
\includegraphics[width=1.0\textwidth]{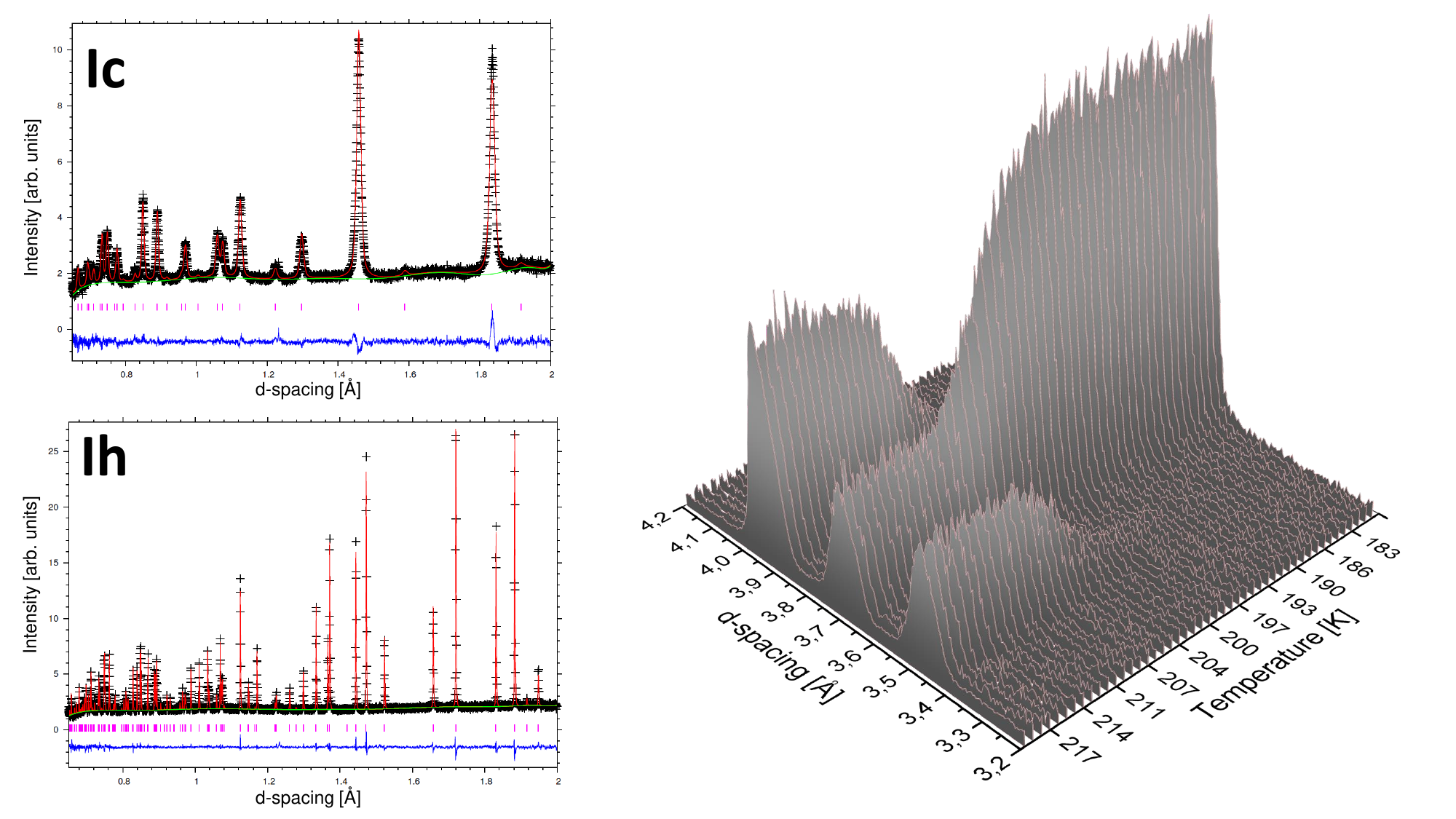}
\caption{\textbf{Representative HRPD diffraction patterns.} Ice Ic (\textbf{top left panel}) and ice Ih (\textbf{bottom left panel}) diffraction patterns collected by the back-scattering bank in the 30–130 ms time-of-flight window and measured at 50 K (black crosses) together with Rietveld fits (red lines) plotted against the interplanar spacing (\textit{d-spacing}). The residuals (blue line) and the positions of the ice Ic (\textit{Fd-3m}) and Ih (\textit{P6$_3$/mmc}) reflections (magenta vertical bars) are shown in each panel. In the right panel a 3D stack-plot contains all the diffraction patterns collected by the forward-scattering bank in the 30–130 ms time-of-flight window and measured during the ice Ic-Ih transformation in the temperature range 180 K $ \leq T \leq $ 220 K, indicated with labels in the right-side axis.
}\label{Diffractograms}
\end{figure}
\begin{figure}[h]
\centering
\includegraphics[width=0.735\textwidth]{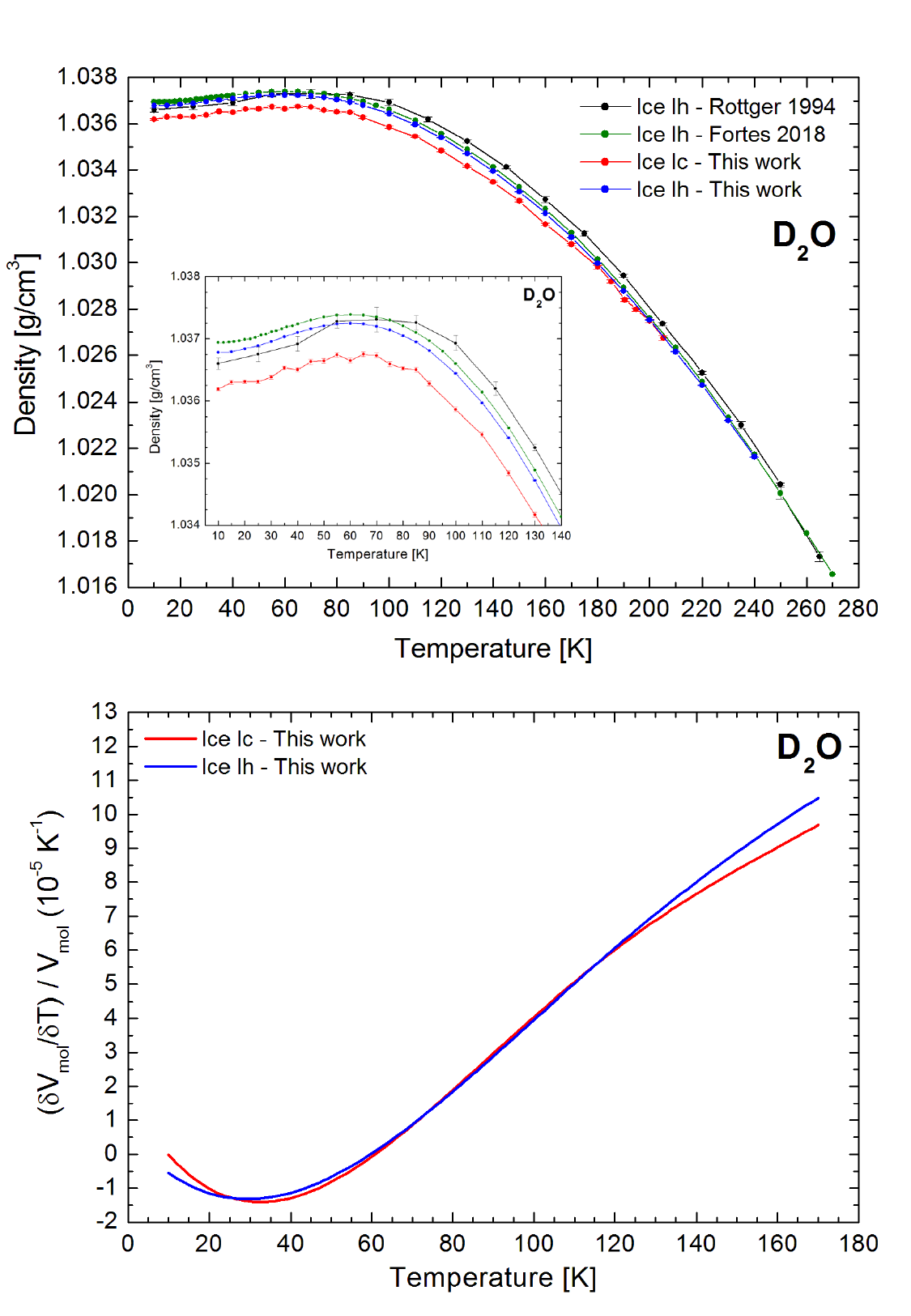}
\caption{\textbf{Density of D$_2$O ice I polytypes as a function of temperature.} (\textbf{Top panel}) Density of D$_2$O samples for ice Ic and ice Ih (red and blue dots, respectively). The values are obtained by means of Rietveld refinement of the diffraction data (back-scattering bank, 30–130 ms time-of-flight window). For comparison, the density of D$_2$O ice Ih as obtained from synchrotron X-ray \cite{Rottger94} and neutron measurements \cite{Fortes18} are reported in the plot, with black and green dots, respectively. (\textbf{Bottom panel}) Linear molecular volume expansivity as a function of temperature for ice Ic (red curve) and ice Ih (blue curve) as calculated from the experimental data.
}\label{Density}
\end{figure}

Two representative HRPD neutron diffractograms, collected by high-resolution back-scattering banks, for both polytypes are reported in Figure \ref{Diffractograms} together with a 3D stackplot of the patterns, collected by the forward scattering bank in the temperature interval 180-220 K, in the d-spacing region containing the the main Bragg reflections for both phases. Here it is possible to clearly appreciate the transformation of ice I from the cubic to the hexagonal symmetry when a slow thermal heating, i.e., 0.2 K/min, is applied. In particular, between 200 K and 210 K, the single reflection (111)$_{CUB}$, peculiar of a pure cubic ice I phase, is replaced by the triplet (101)$_{HEX}$, (002)$_{HEX}$ and (100)$_{HEX}$ proper of hexagonal ice I. The detailed behaviour of the Ih fraction as a function of the temperature is reported in the Fig. SF2 of the Supplementary Information (S.I.).
Knowing the crystal symmetry of the ice Ic and Ih, from the measurements of their lattice parameters obtained by means of a Rietveld refinement of the HRPD data,
we have calculated the density $\rho(T) $ for both the polytypes (see Table \ref{tab_formule}) in a wide range of temperature T using the expression $\rho(T) = m_{\rm D_2 O}/(V_{mol} (T)  N_A) $, where m$_{\rm D_2 O}$ is the mass of a mole of D$_2$O, V$_{mol}$ its molecular volume and $N_A$ is Avogadro's constant. 
\begin{table}[h]
\centering
\caption{Structural parameters relevant for the density calculation of cubic and hexagonal ice I.}\label{tab_formule}%
\begin{tabular}{@{}llllll@{}}
\toprule
Polytype & Space group  & Unit cell symmetry & Molecular volume,  & Neutron measurements\\
 &  & (fitted lattice parameters) &   V$_{mol}$ & temperature range (K)\\
\midrule
Ice Ic    & \textit{Fd}-3\textit{m}  & cubic & $\frac{a^3}{8}$ & 10-205 \\
&  &  (a) &  &  \\
Ice Ih    & \textit{P}6$_3$/\textit{mmc}   & hexagonal & $  \frac{a^2 c \sqrt{3}}{8}$ & 10-240 \\
&  &  (a,c) &  &  \\
\botrule
\end{tabular}
\end{table}
The resulting density is plotted in the upper panel of Figure \ref{Density} together with some reference data taken from literature. In particular, the good agreement between the present determination of the D$_2$O ice Ih density with a previous determination obtained by the reference study of Fortes, reported in Ref. \cite{Fortes18}, allows to confirm the quality of the calibration procedure used for the data reduction from the time-of-flight to the d-spacing values. 
Ice Ic shows a 0.06 \% lower density with respect to ice Ih at base temperature (10 K), with the difference that tends to cancel out as the temperature increases up to its metastability limit. In addition, it is evident the presence of a negative thermal expansion, very similar to that observed in ice Ih, LDA \cite{Eltareb23} and, more generally, in some molecular compounds characterized by a tetrahedral coordination \cite{Evans99}, for which, at low temperature, the shortening of the bond lengths caused by low-energy phonons overcomes the effect of those at higher energy. From a polynomial fit of the experimental density data in the temperature range 10-170 K, it is possible to calculate, in a first approximation, the linear coefficient of the molecular volume thermal expansivity, which is simply given by
\begin{equation}
    \frac{1}{V_{mol} }\frac{\delta V_{mol} }{\delta T} = - \frac{1}{\rho }\frac{\delta \rho }{\delta T}
\end{equation}
The resulting quantity is plotted in the bottom panel of Figure \ref{Density}. After the first negative-expansion region, both curves become positive above about 60 K, even if the position of the minimum is slightly higher in the case of ice Ic (33 K) with respect to ice Ih (29 K). \\ 
\par
The experimental results obtained from the neutron diffraction of ice Ic and Ih were corroborated by means of constant-volume ($NVT$) molecular dynamics simulations in which the molar enthalpy [i.e., $H_c(T)$ and $H_h(T)$] of these two structures was calculated as a function of temperature in the range 60.0 K $ \leq T \leq $ 205.3 K. 
Although many details of these simulations are reported in Methods section, at this stage it is important to emphasize two aspects of molecular dynamics calculations: the choice of intermolecular potential and the inclusion of quantum effects in the evaluation of 
$H_c(T)$ and $H_h(T)$.
\begin{figure}[h]
\centering
\includegraphics[width=1.0\textwidth]{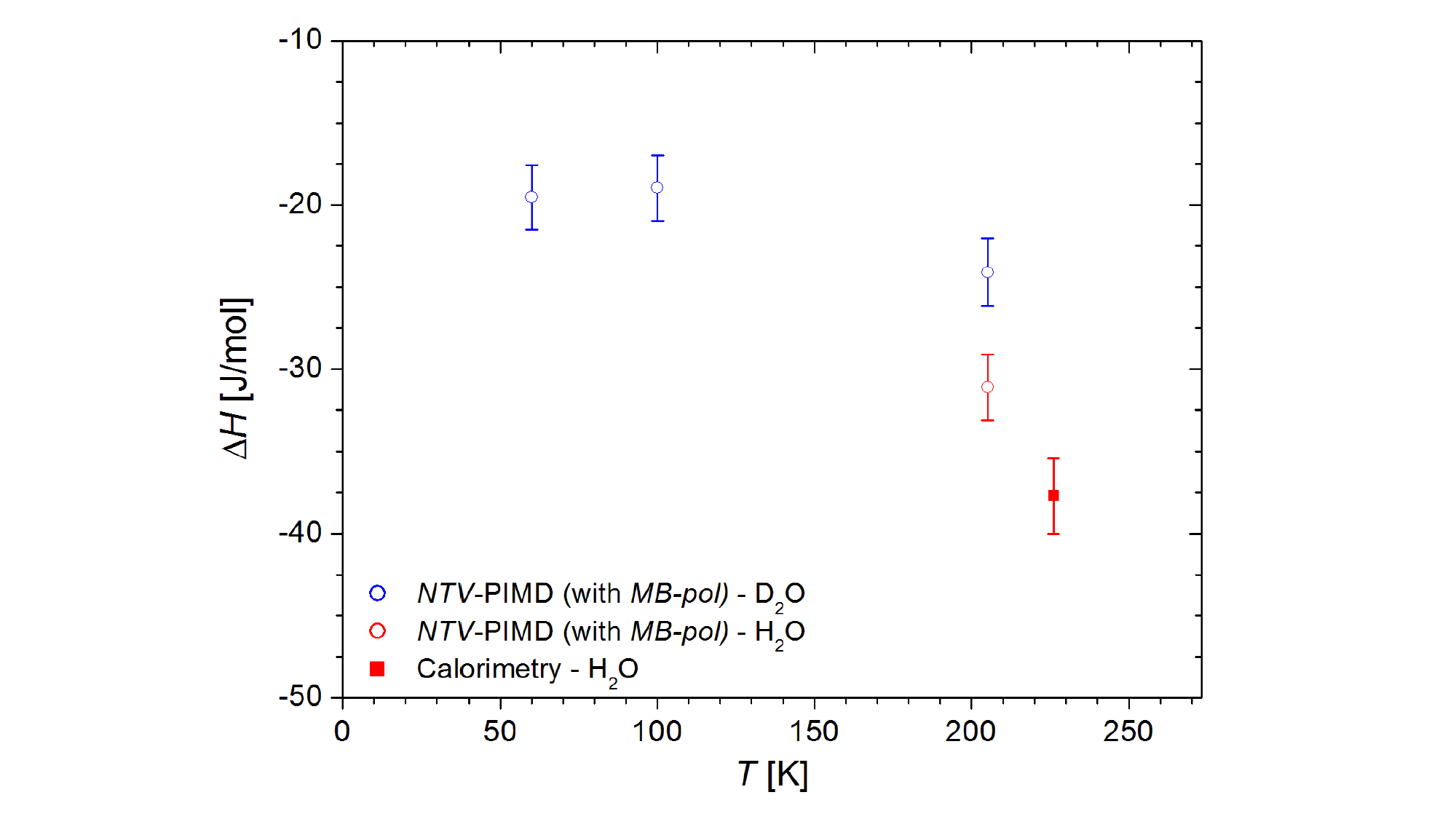}
\caption{\textbf{Enthalpy jump $\Delta H(T)= H_h(T)-H_c(T)$ for the two ice I polytypes as a function of temperature.} Blue empty circles with error bars stand for the present molecular dynamics simulation data on D$_2$O, red empty circle with error bar on H$_2$O, while the red full square with error bar represents the H$_2$O calorimetric measurement from Ref. \cite{Tonauer23bis}.
}
\label{PIMC}
\end{figure}
As for the former, we have resorted to the well-known {\it MB-pol(2023)} many-body potential \cite{Paesani23}, while dealing with the latter we have implemented a standard version of the Path Integral Molecular Dynamics (PIMD) algorithm \cite{Ceriotti10}, where nuclear quantum effects are taken into account via Feynman path integrals.
It is worth noting that the joint use of {\it MB-pol(2023)} and PIMD allows to obtain state-of-the-art simulations of the thermodynamic properties of H$_2$O and D$_2$O in a large portion of their phase diagrams, in particular in the temperature range from 200 to 300 K as clearly shown in Ref. \cite{Bore23}. 
In this way, it was possible to evaluate the enthalpy jump $\Delta H(T)= H_h(T)-H_c(T)$, which for $T=205.286$ K was estimated to be $(-24.1 \pm 2.1)$ J/mol for D$_2$O and $(-31.1 \pm 2.0)$ J/mol for H$_2$O, the latter being in good agreement with the most recent experimental finding obtained via calorimetric measurements with hydrogenated ice \cite{Tonauer23bis}, that is $\Delta H(T^\ast)= (-37.7 \pm 2.3)$ J/mol at $T^\ast \approx 226$ K.
The full trend of $\Delta H(T)$ is plotted in Figure \ref{PIMC} in the mentioned temperature range.\\
\par
\textbf{In conclusion}, this study allowed to measure the density of cubic ice in its entire thermodynamic range of existence, finding a direct correspondence between the small enthalpy difference between the two ice I polytypes and their crystalline structures. 
As a matter of fact, there is no temperature at which the ice Ic has the lower enthalpy.
At this stage, we have no access to the corresponding entropy changes, $\Delta S$, needed to fully assess the Gibbs free energy difference, $\Delta G$, between these two ice I polytypes for $T \le 226$ K. 
The latter physical quantity has been evaluated for H$_2$O only at $T=253$ K, amounting to $\Delta G=(-16.5 \pm 1.7)$ J/mol \cite{hondoh15,Lupi17}, and allows just for a very crude estimate of the entropy jump in the $(226-253)$ K temperature interval, once considerered the deutereted polytypes, that is: $\Delta S_{\rm D_2O}= S_{h,\rm D_2O} - S_{c,\rm D_2O} \approx -0.10$ J/K/mol.
However, assuming the reasonable hypothesis that $ S(T)$, which is essentially due to phonons since the residual Pauling entropy is practically identical in the two cases \cite{Paul24}, monotonically decreases for $T$ getting lower than about 230 K--250 K, we obtain
\begin{equation}
    \Delta H_{\rm D_2O} < T \Delta S_{\rm D_2O} \text{   with    } T < 205 \text{ K}.
\end{equation}
Thus, all the present quantum calculations, based on measured densities, quantitatively demonstrate the metastable nature of Ic with respect to the other polytype (i.e., ice Ih), even in the low temperature regime.
In addition to the paramount importance for the ice thermodynamics itself and for the physics of the atmosphere, with the opening of the possibility for the remote sensing of some of the ice properties like density and cubicity by means of advanced Earth orbit instrumentation like the James Webb Space Telescope (JWST) \cite{Tonauer24}, this study can be of interest also in the astrophysical context, where low temperatures and high vacuum conditions are a plausible scenario, especially on the icy crusts of airless small objects like both natural satellites and asteroids, e.g., Europa and Ceres, respectively.
\section{Methods}
\label{Methods}
\subsection{Sample synthesis and Raman characterization}
Pure D$_2$O ice, in the form of fine powder, was inserted in a Cu-Be high-pressure container, kept at 256 K, and pressurized at 4.3 kbar with H$_2$ gas, thus obtaining about 20 cm$^3$ of hydrogen filled ice in the C$_0$ phase \cite{delRosso16}. The sample was then recovered at ambient pressure by quenching it at liquid nitrogen temperature (i.e., 77 K) and transferred, without interrupting the cold chain and avoiding air condensation onto the sample surface, in a large Raman optical cell of about 15 cm$^3$, suitable for thermal treatment of a large amount of sample, and monitored via Raman spectroscopy (see Supplementary Note 1 in the S.I.). In particular, after a first annealing at 120 K under dynamic vacuum, the resulting emptied hydrate, i.e., D$_2$O ice XVII, was heated, at constant rate of 0.04 K/min and in the same vacuum conditions, up to 160 K, applying then an isothermal waiting of about 2 hours to ensure that all the sample was transformed in pure cubic ice Ic (Figure SF1 of the S.I.). Then the ice Ic sample was cooled again to 77 K, extracted from the optical cell and recovered in cryogenic vials, which were stored in a dry nitrogen Dewar suitable for shipping to ISIS facility for neutron diffraction experiment.
\subsection{Neutron diffraction}
The neutron diffraction experiment was performed with the HRPD neutron time-of-flight diffractometer at ISIS (Rutherford Appleton Laboratory, U.K.) \cite{Ibberson09}. The ice powder sample was loaded in a slab vanadium cell, with an internal thickness of 10 mm, and placed in the HRPD sample tank, where a closed-cycle refrigerator allows to set the temperature in the range 10-240 K. For each thermodynamic point, usually set with a 10 K step change, we waited 15 min for the effective sample thermal equilibration before starting the measurement, with the exception of the diffractograms collected close to the Ic-Ih transformation (180-220 K), when a warming ramp at the constant rate of 0.2 K/min was applied. Further details about the thermodynamic protocol applied in the diffraction experiment are reported in Supplementary Note 2 of the S.I. \\
\\
Data, available at the ISIS database (see Supplementary Note 3 of the S.I.), were mainly collected in the 30-130 ms time-of-flight window, covering 0.65-10.2 \AA  with the three detector banks, i.e., back-scattering ($2\theta=168.3 \deg$), equatorial ($2\theta=90 \deg$) and forward ($2\theta=30 \deg$). Given the tiny difference between the two ice polytypes in terms of density to be measured (that is, below 0.1 \%) we needed to collect at least an integrated proton current of 15-20 $\mu$A for each thermodynamic point set, in order to have good statistics in the best resolution diffractograms (i.e., $\Delta d / d \simeq 6 \times 10^{-4}$ with the back-scattering bank), that are those refined to extract the density values. By means of Mantid routine, 
raw data were time-focused, normalized to incident spectrum and corrected for instrumental efficiency. The reduced data were then refined by Rietveld method using the GSAS software package. 
A new instrument parameter file, necessary to convert time-of-flight to d-spacing and to determine the peak profile constants, was built using the diffraction, measured at the beginning of the facility cycle, of the NIST silicon standard SRM 640e and of the CeO$_2$, respectively. The diffraction data were fitted using a peak profile function 3 (i.e., a convolution of back-to-back exponentials with a pseudo-Voigt function). The instrumental broadening parameters $\alpha-0$, $\beta-0$ and $\beta-1$, determined by previous refinement of the CeO$_2$ data, were kept fixed during the refinement of the ice datasets, while the other profile parameters $\sigma-1$ and $\gamma-1$ (Gaussian and Lorentzian, respectively), which model the peak broadening caused by the sample, were always refined. Further details about the refined parameters are reported in the Supplementary Note 3 of the S.I..
\subsection{Path-Integral Molecular Dynamics Simulations}
In order to rigorously capture nuclear quantum effects, PIMD simulations, on both H$_2$O and D$_2$O, were performed using the i-PI universal force engine \cite{Ceriotti10} coupled with the MBX software \cite{Paesani23} for the evaluation of the many-body potential energy and forces in water. 
The equations of motion were integrated in the canonical ($NVT$) ensemble using a local path integral Langevin equation (PILE-L) thermostat \cite{Ceriotti10} with a global time constant $\tau = 100$ fs and a time step $\Delta t = 0.25$ fs.
To ensure a uniform convergence of the quantum properties, the discretization of the Feynman path integral was linearly adapted to the system temperature: $PT \approx \text{const.}$, with $P$ being the number of beads (also known as Trotter number). 
In this way, we employed $P=109$  at $T=60$ K, $P=64$ at $T=100$ K, and $P=32$ at $T=205.286$ K for both cubic (Ic) and hexagonal (Ih) ice phases.
Short-range interactions were evaluated within a cutoff of 9.0 \AA{} and 4.5 \AA{} for the two-body terms and the three-body terms, respectively.
Long-range electrostatic and dispersion interactions were computed using the particle-mesh Ewald (PME) summation with a grid density of 2.5~\AA$^{-1}$, a spline order of 6, and an Ewald splitting parameter $\alpha = 0.60$.
The induced dipoles were converged using the conjugate gradient method with a tolerance of $10^{-10}$ hartree in order to ensure a rigorous energy conservation.
Each system was equilibrated for a time of 100 ps, followed by a production run of 100 ps for the actual data collection.
\subsection{Thermodynamic Analysis}
The enthalpy $H(T)$ of each ice sample was computed from the ensemble-averaged energies extracted from the production trajectories.
It was defined as:
\begin{equation}
H(T) = \langle U \rangle_{NVT} + \langle K \rangle_{NVT} + p_{\text{exp}}V_{\text{m}}
\end{equation}
where $\langle U \rangle_{NVT}$ and $\langle K \rangle_{NVT}$ represent the average quantum potential and kinetic energies, respectively. 
The so-called pressure-energy term $p_{\text{exp}}V_{\text{m}}$ was calculated using the experimental pressure value $p_{\text{exp}}$ and the diffraction molar volume $V_{\text{m}}$ corresponding to the fixed density of the simulation cell.
The relative stability of the phases was assessed by calculating the enthalpy difference $\Delta H(T) = H_{h}(T) - H_{h}(T)$ at each temperature value.
This methodology yielded thermodynamic differences that are robust against the virial pressure fluctuations of PIMD, and consistent with the experimental thermochemical data.
\backmatter
\bmhead{Supplementary Information}
Supplementary Note 1-3, Figure 1-2, Table 1-2.
\bmhead{Acknowledgements}
We would like to thank A. Donati (CNR-IFAC) for his valuable and fundamental  technical contribution, and L. Ulivi (CNR-IFAC) for his scientific contribution.
We also thank ISIS for providing neutron beam time and the ISIS sample environment staff for their fundamental technical support. 
\section*{Declarations}
\begin{itemize}
\item \textbf{Funding}

This research has been funded by the European Union - NextGeneration EU, within PRIN 2022, PNRR M4C2, Project “E-ICES” 2022NRBLPT\_PE3\_PRIN2022 (CUP: B53D23004390006).  We also acknowledge the financial support of the DIITET-CNR with the project "STRIVE" (DIT.AD022.207, CUP: B53C22010110001). This research was also funded by the European Union – Next Generation EU from the Italian Ministry of Environment and Energy Security POR H2 AdP MMES/ENEA with involvement of CNR and RSE, PNRR - Mission 2, Component 2, Investment 3.5 "Ricerca e sviluppo sull’idrogeno" (PRR.AP015.017 - Ricerca e sviluppo di tecnologie per la filiera dell'idrogeno H2 - AdC ENEA/CNR POR IDROGENO, CUP: B93C22000630006).
\\
\item \textbf{Conflict of interest}

The authors have no conflicts to disclose.
\\
%
%
%
%
\item \textbf{Data availability} 

The data supporting the findings of the present work are available from the corresponding authors upon reasonable request.
\\
\item \textbf{Author contribution}

This work is the result of a common effort to which all authors contributed. In particular: L.d.R conceived and supervised the study. L.d.R. and D.C. acquired the fundings. M.C. managed the resources. L.d.R., M.C. and D.C. designed the experiments and the simulations. L.d.R. and M.C. set up the high pressure apparatus and synthesized the samples. L.d.R. and M.C. carried out the Raman experiments and the initial sample treatment. A.D.F. carried out the experiment at ISIS, RAL. L.d.R. and M.C. performed the Raman data analysis. L.d.R., F.G. and S.B. performed the diffraction data analysis. A.S. and D.C. carried out the molecular dynamic simulations and wrote the corresponding part of the manuscript. L.d.R. wrote the manuscript with original contributions from D.C., A.S. and S.B.  All the authors read and reviewed the manuscript.
\\
\end{itemize}
\bibliography{Density_iceIc}
\end{document}